\documentclass[twocolumn,letterpaper,10pt,amsmath,amssymb,superscriptaddress]{revtex4}

\pdfoutput=1

\usepackage{graphicx}
\usepackage{dcolumn}
\usepackage{bm}
\usepackage{ulem}
\usepackage{color}
\usepackage{amsmath}
\usepackage[sort&compress]{natbib}

\def\H{\mathbf{H}}
\def\b{\mathbf{b}}
\def\bdag{\mathbf{b}^{\dagger}}
\def\sigm{\mathbf{\sigma}_-}
\def\sigp{\mathbf{\sigma}_+}

\def\wqd{\omega_{qd}}
\def\wcav{\omega_{cav}}
\def\wcon{\omega_{c}}
\def\kin{\kappa_{||}}
\def\ein{\epsilon_{in}}
\def\L{\mathbf{L}}
\def\Imax{I_{\max}}
\def\Imin{I_{\min}}
\def\gqd{\gamma_{qd}}
\def\Pwg{P_{wg}}
\def\Pinc{P_{inc}}
\def\C{\mathbf{C}}
\def\D{\mathbf{D}}

\begin{document}

\title{Attojoule all-optical switching with a single quantum dot}

\author{Deepak Sridharan}
\affiliation{Department of Electrical and Computer Engineering, IREAP, University of Maryland, College Park,
Maryland 20742, USA and Joint Quantum Institute, University of Maryland, College Park, Maryland 20742, USA}

\author{Ranojoy Bose}
\affiliation{Department of Electrical and Computer Engineering, IREAP, University of Maryland, College Park,
Maryland 20742, USA and Joint Quantum Institute, University of Maryland, College Park, Maryland 20742, USA}

\author{Hyochul Kim}
\affiliation{Department of Electrical and Computer Engineering, IREAP, University of Maryland, College Park,
Maryland 20742, USA and Joint Quantum Institute, University of Maryland, College Park, Maryland 20742, USA}

\author{Glenn S. Solomon}
\affiliation{Joint Quantum Institute, National Institute of Standards and Technology, and
University of Maryland, Gaithersburg, Maryland 20899, USA}

\author{Edo Waks} \email{edowaks@umd.edu}
\affiliation{Department of Electrical and Computer Engineering, IREAP, University of Maryland, College Park,
Maryland 20742, USA and Joint Quantum Institute, University of Maryland, College Park, Maryland 20742, USA}

\date{\today}

\begin{abstract}
We experimentally investigate the dynamic nonlinear response of a single quantum dot (QD)
strongly coupled to a photonic crystal cavity-waveguide structure. The temporal response
is measured by pump-probe excitation where a control pulse propagating through the
waveguide is used to create an optical Stark shift on the QD, resulting in a large
modification of the cavity reflectivity. This optically induced cavity reflectivity
modification switches the propagation direction of a detuned signal pulse. Using this
device we demonstrate all-optical switching with only 14 attojoules of control pulse
energy. The response time of the switch is measured to be up to 8.4 GHz , which is
primarily limited by the cavity-QD interaction strength.
\end{abstract}

\maketitle


All-optical switches are considered to be an important alternative for increasing
information bandwidth and reducing power consumption in telecommunications systems and
computer processors~\cite{Miller2010}.  Optical switching has been demonstrated using
various device structures such as semiconductor quantum
wells~\cite{GopalYoshida2002,TakahashiKawamura1996}, semiconductor optical
amplifiers~\cite{NakamuraUeno2001,DorrenYang2004,NielsenMork2006}, and nonlinear
parametric processes~\cite{AndreksonSnnerud2008}.  These devices typically exploit weak
nonlinearities arising from a large ensemble of atomic systems, resulting in high power
dissipation and large device size~\cite{HintonRaskutti2008}.  Photonic crystals (PCs)
have been shown as an effective method for significantly reducing both device size and
power consumption. PC all-optical switches have been demonstrated using free carrier
absorption~\cite{HuskoRossi2009,NozakiTanabe2010} and laser gain
modulation~\cite{MatsuoShinya2010}, enabling switching energies as low as 0.6-15
femtojoules.

%

The interaction between PCs and semiconductor quantum dots (QDs) provides a promising
method for achieving significantly enhanced nonlinear optical response.  These
interactions can be sufficiently large to enter the strong coupling regime of cavity
quantum electrodynamics, where atom light interactions modify both the QD emission
spectrum~\cite{YoshieScherer2004} and cavity
spectrum~\cite{HughesKamada2004,WaksVuckovicPRL2006,EnglundFaraon2007,ShenFan2009,BoseSridharanOpEx2011}.
Such modifications can result in nonlinear optical effects near the single photon level,
which has been predicted theoretically~\cite{WaksVuckovicPRA2006,AuffevesSimon2007} and
reported
experimentally~\cite{SrinivasanPainter2007,FushmanEnglund2008,BoseSridharanAPL2011} in a
number of works. The dynamics of the nonlinear response of a strongly coupled cavity-QD
system remain largely unexplored to date.  A better scientific understanding of this
dynamic nonlinear behavior could provide important insight for application of these
systems for both classical and quantum information processing.


\begin{figure}
\centering
\includegraphics[width = 3.0in]{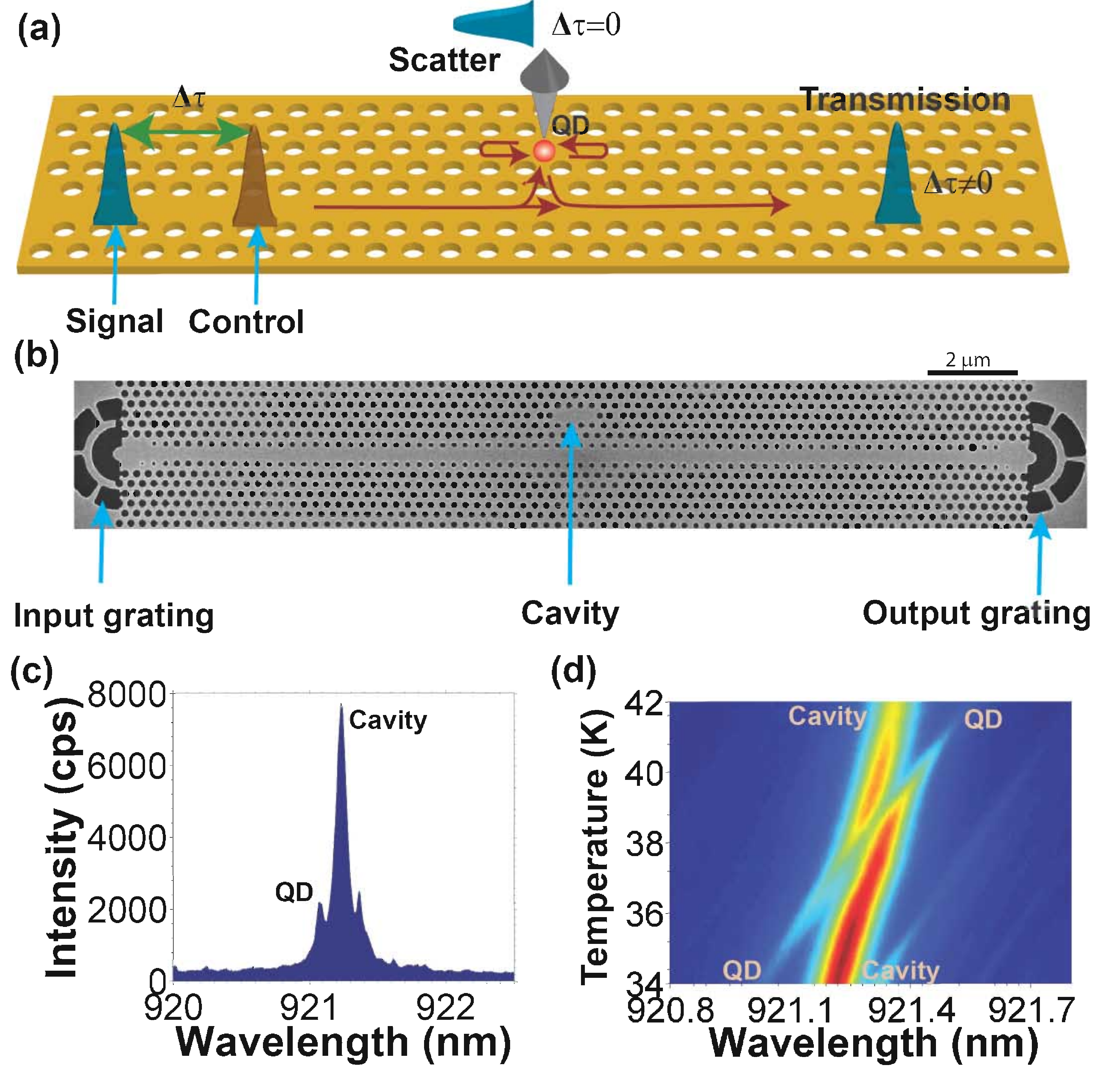}
\caption{(a) Schematic of cavity-waveguide system. (b) Scanning electron micrograph showing the
fabricated device showing cavity, input grating, and output grating (c) Low power PL measurement of cavity.
The QD studied in this letter is labeled.  (d) Cavity
PL as a function of sample temperature.} \label{fig:Device}
\end{figure}

In this work we experimentally study the dynamic nonlinear response of a QD strongly
coupled to photonic crystal cavity-waveguide circuit using optical pump-probe
measurements. Figure 1a illustrates the device, which is composed of a photonic crystal
defect cavity evanescently coupled to a row-defect waveguide, with a single QD strongly
coupled to the cavity mode. The pump-probe experiment proceeds by injecting a signal and
control beam, which are spectrally detuned, into the waveguide.  The control beam
determines whether the signal will be preferentially transmitted through the waveguide or
scattered by the cavity.  A scanning electron micrograph (SEM) image of a fabricated
device is shown in Figure 1b. Details of the device design and fabrication have been
previously reported~\cite{BoseSridharanOpEx2011}. The signal and control pulses are
injected into the waveguide using a grating coupler~\cite{FaraonFushman2008}, and drive
the cavity evanescently. The signal pulse is collected either directly from the cavity
(direct cavity scatter) or from the output coupler (transmitted waveguide signal) by
spatial filtering.

Figure 1c shows the cavity photoluminescence (PL) spectrum, attained by exciting the
cavity with a 780 nm pump laser.  The PL exhibits an emission peak for the cavity mode,
along with additional emission peaks for several QDs that are coupled to the cavity. The
QD used for all measurements reported in this letter is labeled in the figure. By fitting
the cavity mode to a Lorentzian, we determine the cavity energy decay rate to be
$\kappa=28.0$ GHz (Q~11900). Figure 1d shows the photoluminescence as a function of
device temperature. As the temperature is increased, the QD identified in Figure 1c
red-shifts and becomes resonant with the cavity.  A clear anti-crossing is observed,
indicating that the QD and cavity are in the strong coupling regime and form two dressed
polariton modes. From the minimum splitting of the polaritons, which occurs at a
temperature of 39 K, we calculate the cavity-QD coupling strength to be $g/2\pi =13.4$
GHz.

\begin{figure}
\centering
\includegraphics[width = 3.0in]{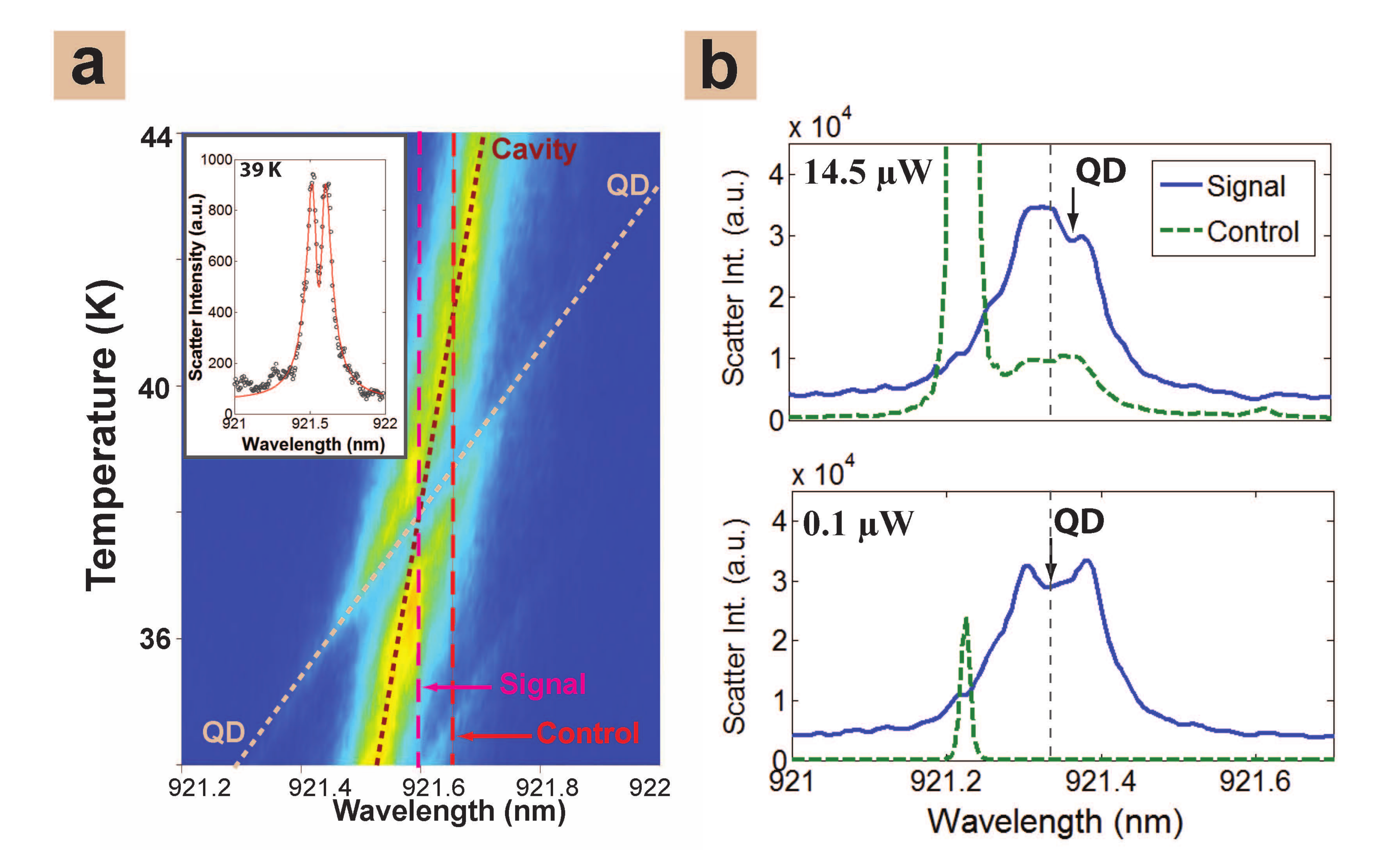}
\caption{(a) Cavity scatter under broadband LED excitation as a function of temperature.
Dotted lines indicate the temperature dependence of QD and cavity.  Dashed lines indicate
the frequencies of the signal and control pulses in the pulsed experiments.  The inset
shows a single spectrum taken at 39 K when the QD is resonant with the cavity.  (b)
Cavity scattering spectrum for two different control field powers. The dashed green line
shows the cavity scattering spectrum with only the control field (no signal).  The solid
blue line shows the scattering spectrum of only the signal.}\label{fig:CW}
\end{figure}

In Figure 2a we plot the measured cavity scattering spectrum as a function of device
temperature when a broadband LED, used as a white light source, excites the input
grating.  Similar to the PL measurements, an anti-crossing between QD and cavity mode is
observed in the resonant scattering spectrum of the cavity.  At the QD resonant frequency
the cavity scattering is suppressed due to strong
coupling~\cite{HughesKamada2004,WaksVuckovicPRL2006,EnglundFaraon2007,ShenFan2009}. The
inset shows the measured cavity scattering spectrum taken at 39 K when the QD is resonant
with the cavity, which exhibits a doublet due to strong cavity-QD interactions.  The
solid line in the inset is a theoretical fit assuming a Jaynes-Cummings interaction
model~\cite{BoseSridharanOpEx2011,WaksSridharan2010}.

We first investigate the nonlinearity of the device under continuous wave excitation by
injecting a second control field from a narrowband tunable external cavity diode into the
input grating along with the broadband LED.  The control field is detuned from the cavity
resonance by 0.12 nm.  Figure 2b shows the resulting scattering spectrum, taken when the
QD is resonant with the cavity, for two different control field powers.  The dashed green
line shows the scattering spectrum when only the control is present.  At a control power
of 14.5~$\mu$W (measured before the input grating) we observe indirect scatter from the
cavity polaritons due to non-resonant energy
transfer~\cite{WingerVolz2009,AtesUlrich2009,EnglundMajumdar2010,MajumdarFaraon2010}. The
blue curve shows the scattering spectrum of the broadband LED when injected with the
control, where we have subtracted the indirect scatter from the control field.  At
0.1~$\mu$W of control power the control field is weak and does not affect the cavity
scatter, which exhibits a dip at the QD resonant frequency.  As the control field power
is increased to 14.5 $\mu$W, we see that the position of the dip induced by the QD is red
shifted due to the optical Stark effect and the contrast is slightly reduced due to
saturation~\cite{FushmanEnglund2008}. The Stark shift enables the control field to
optically modify the amount of signal scattered or transmitted, providing the possibility
for all-optical switching. We note that the contrast of the dip in Figure 2b is reduced
as compared to the inset to Figure 2a, even at low control powers. This reduction occurs
because we are using a relatively large signal power to minimize the contribution of
incoherent photons from the control, which partially saturates the QD.

We next perform dynamic pump-probe measurements where the control and signal pulses are
generated by two synchronized Ti:Sapphire lasers.  The pump laser has a pulse duration of
80 ps.  The probe laser initially emits 5 ps pulses that are filtered down to a bandwidth
of 0.02 nm (7 GHz), corresponding to a 60 ps pulse duration, using a fiber Bragg grating.
The bandwidth of the probe laser is chosen to be approximately half the spectral width of
the dip induced by the QD in the scattering spectrum, which is equal to $g$ in the strong
coupling regime.  The pump pulse train is synchronized to the probe by a piezo feedback
in the pump laser cavity, and the delay between the pump and probe is controlled
electrically by a phase-locked loop in the synchronization circuit. Measurements are
taken at a fixed signal and control frequency, which are selected such that when the QD
is resonant with the cavity, the signal is resonant with the QD and the control is
resonant with the lower polariton as indicated by the dashed lines in Figure 2a.  The
signal field intensity is set to be sufficiently weak to be in the linear response regime
of the cavity-QD system. Temperature tuning is used to tune the QD through the signal
field center frequency. The signal and control fields can be collected either directly
from the cavity or from the output coupler, and are separated by a grating spectrometer.

\begin{figure}
\centering
\includegraphics[width = 3.0in]{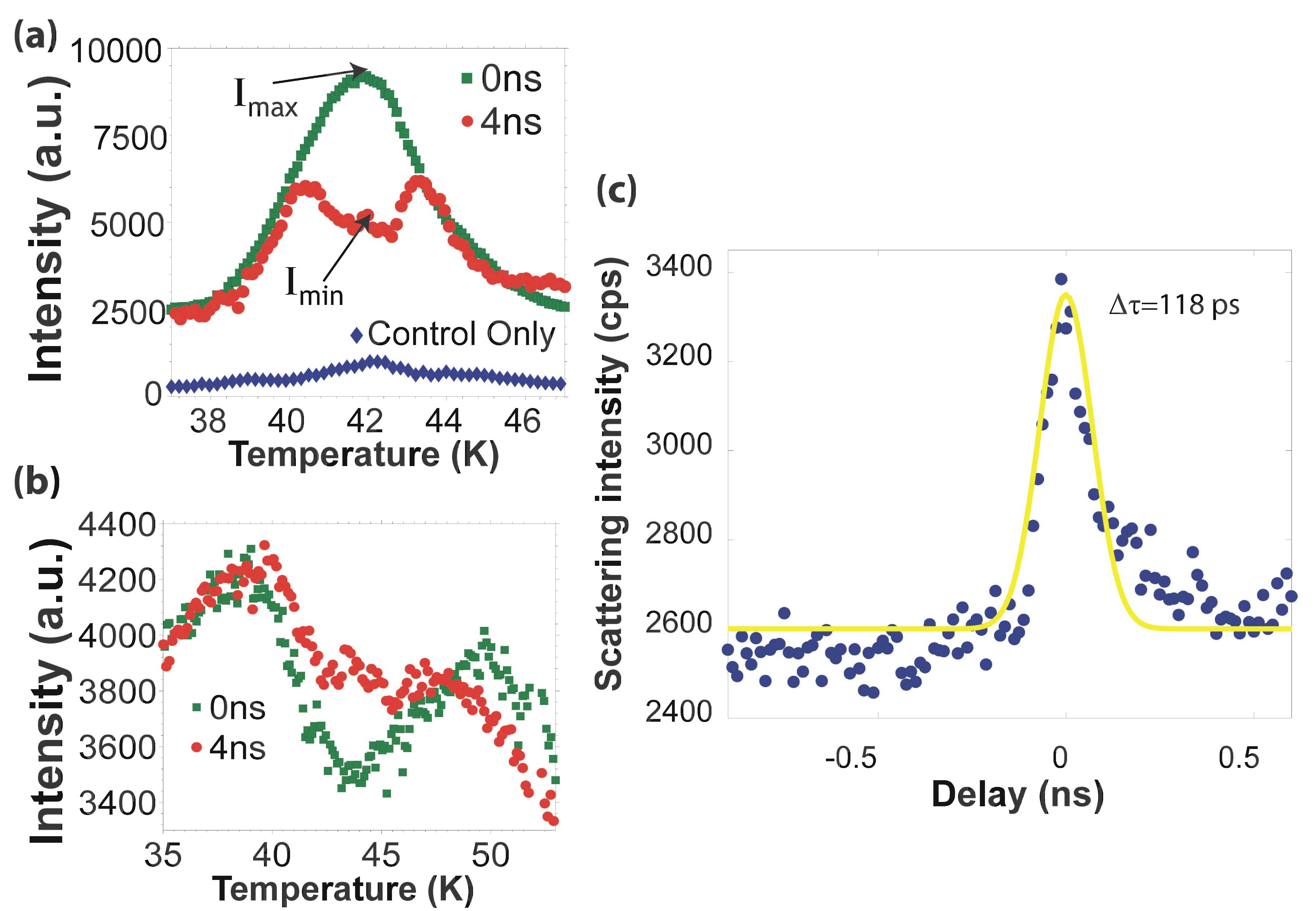}
\caption{The signal scattering intensity as a function of sample temperature for delays of 0 ns
(green squares) and 4 ns (red circles) is shown (a) at the cavity radiation port and (b)
at the transmission port (output coupler).  Scattering spectrum with control signal only
indicated as blue diamonds in panel a.  (c) Cavity scattering intensity at the strong
coupling point as a function of delay between the signal and control pulses. The solid
line represents a Gaussian fit to the data.}\label{fig:Pulsed_Time}
\end{figure}

Figure 3a shows the scattering intensity collected directly from the cavity as a function
of temperature when the delay between signal and control is set to both 0 ns and 4 ns. At
4 ns delay, the pulses excite the cavity at different times and therefore do not
interact. In this case the cavity scattering spectrum is suppressed when the QD is
resonant with the signal energy (41.8 K) due to cavity-QD interactions. We note that the
temperature where resonance is achieved is slightly different than that of Figure~2a
because of a gradual red-shift of the cavity resonance frequency which is observed over
the course of the measurement process.  When the control and signal arrive simultaneously
at the cavity (zero ns delay), we observe a significant increase of the cavity scatter at
same temperature.  The switching contrast, defined as $\delta=(\Imax-\Imin)/\Imax$ where
$\Imax$ and $\Imin$ are the scattering intensities at the QD resonant frequencies at zero
and large delays respectively, is calculated to be 0.44. Figure 3b shows the behavior of
the signal transmitted to the output coupler, which exhibits the conjugate effect where
transmission is enhanced when the QD is resonant with the signal frequency.

In Figure 3a we also plot the case where only a control pulse is injected (blue
diamonds).  In this case there is still some optical energy at the QD frequency. The
noise to signal ratio is 0.18 (at 0 ns delay), which is lower than the value of 0.35 for
the CW case shown in the top panel of Figure 2b, even though in pulsed operation the
signal is well below QD saturation and the detuning between the QD and control frequency
is significantly smaller.  Furthermore, in the pulsed case the noise is mostly dominated
by the spectral overlap between the signal and control as opposed to non-resonant energy
transfer.  Thus, although noise injection from the control field can be a problem in CW
measurements, it is a much smaller effect for pulsed switching operation.  We attribute
this difference to the fact that incoherent control field scattering is proportional to
the average control power while Stark shift and saturation are proportional to peak
control power.  In pulsed operation, we can achieve a high peak power with a relatively
low average power, which significantly reduces incoherent scattering. To measure the
switching speed of the system, we fix the sample temperature at the strong coupling point
and plot the cavity scatter as a function of delay between pump and probe, as shown in
Figure 3c.  The scatter exhibits a sharp peak near 0 ns delay, which is fit to a Gaussian
function.  From the fit we determine that the switching occurs over a 118 ps response
time.

To determine the control pulse energy, we need to know the efficiency with which light is
injected into the waveguide mode by the grating coupler.  We use an optical Stark shift
measurement to precisely measure this coupling efficiency~\cite{BoseSridharanAPL2011}.
The measurement proceeds by setting the sample temperature to 45 K where the QD is red
shifted from the cavity mode by 55 GHz. A tunable external cavity laser diode is focused
onto the input grating and tuned to be resonant with the cavity mode.  We monitor the QD
emission through non-resonant energy transfer and record the spectrum as a function of
pump power. For each spectrum the emission from the QD is fit to a Lorentzian function to
determine the center frequency. The experimental result of Stark shift of the QD as a
function of pump power are plotted as red circles in Figure 4.

\begin{figure}
\centering
\includegraphics[width = 2.0in]{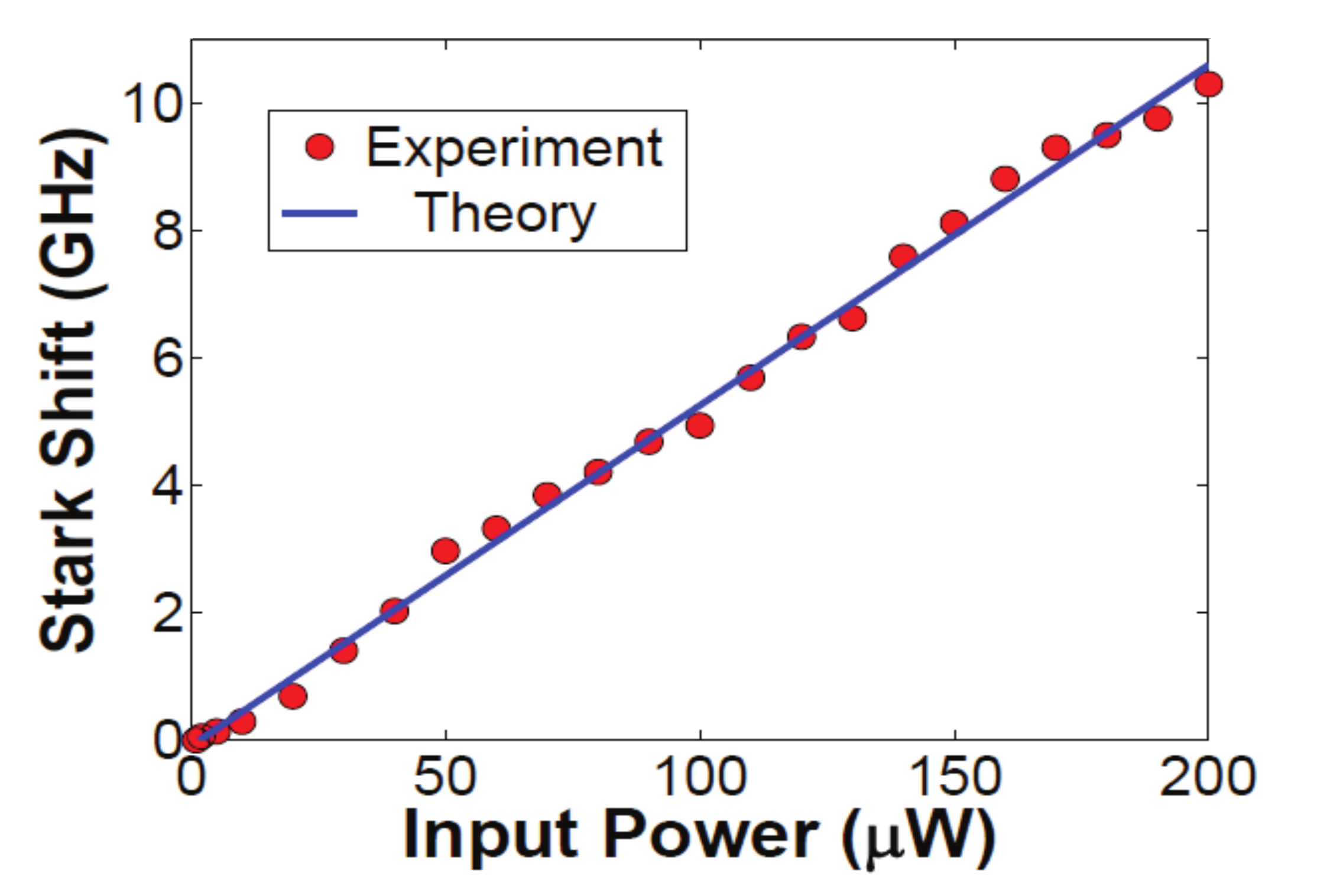}
\caption{Stark shift of the QD versus input power at the cavity mode resonance (red circles) along
with the fit from numerical simulations (solid blue line). }\label{fig:Stark}
\end{figure}

The Stark shift data can be directly used to determine the energy propagating in the
waveguide by fitting the results to numerical simulations performed using the full master
equation formalism. The master equation is given by $\dot{\rho} = [\rho,\H]/i\hbar +
\L\rho$ where $\rho$ is the density matrix of the cavity-QD system, $\mathbf{H}$ is the
Hamiltonian, and $\mathbf{L}$ is the Liouvillian superoperator that accounts for losses
and decay of the system.  We define $\b$ as the bosonic annihilation operator for the
cavity field, and $\sigm$ as the dipole lowering operator for the QD. The Hamiltonian is
given, in the reference frame of the input field, by $\H = \hbar\Delta \sigp\sigm
+\sqrt{\kin}\ein\left( \bdag+\b \right) +\hbar g (\bdag\sigm +\sigp\b)$ where $\Delta$ is
the detuning between the cavity and QD, $\kin$ is the in-plane coupling rate between the
cavity and the QD, and $\ein$ is the electromagnetic field amplitude traveling in the
waveguide.  The Liouvillian is given by $\L = \kappa/2 \D(\b) + \gamma/2 \D(\sigm) +
\gamma_d \D(\sigp\sigm)$ where for any collapse operator $\C$ we have $\D(\C) =
2\C\rho\C^\dagger - \C^\dagger\C\rho - \rho\C^\dagger\C$.  We treat the coupling
efficiency   as fitting parameter given by  $\eta=\Pwg/\Pinc$, where $\Pinc$ is the
experimentally measured incident power and  $\Pwg = \hbar\wcav\ein^2$ is the optical
power propagating in the waveguide.

Simulations are performed using the measured values of $g$ and $\kappa$.  The QD decay
rate, obtained from a numerical fit of the inset to Fig. 2a, is given by $\gqd/2\pi =
5.8$ GHz.  To determine the in-plane coupling rate $\kin$, we perform in-plane
transmission measurements of the waveguide at a temperature of $51$~K so that the QD is
well detuned from the cavity mode. The measurement procedure for obtaining this decay
rate is described in detail in Ref.~\cite{BoseSridharanOpEx2011}, and the detailed
measurement results for the device used in this work are provided in the supplementary
material. From these measurements we obtain the in-plane decay rate of $\kin/2\pi = 2.9$
GHz. The power spectrum of the cavity mode, defined as the Fourier transform of the
two-time covariance function $F(\tau)=\langle \bdag(t+\tau)\b(t)\rangle$, is calculated
as a function of $\Pwg$ using standard quantum regression theory. From the power spectrum
we determine the center wavelength and linewidth of the QD using a Lorentzian fit, and
numerically optimize $\eta$ to achieve best agreement between simulation and experimental
results based on the measured incident power $\Pinc$. All calculations are performed
using an open source quantum optics toolbox~\cite{Sze1999}. The waveguide coupling
efficiency is found to be $\eta = 1.4\times10^{-3}$. The solid line in Fig.~4 shows the
theoretical fit using this coupling efficiency, which exhibits superb agreement with the
experimentally measured values.

\begin{figure}
\centering
\includegraphics[width = 3.0in]{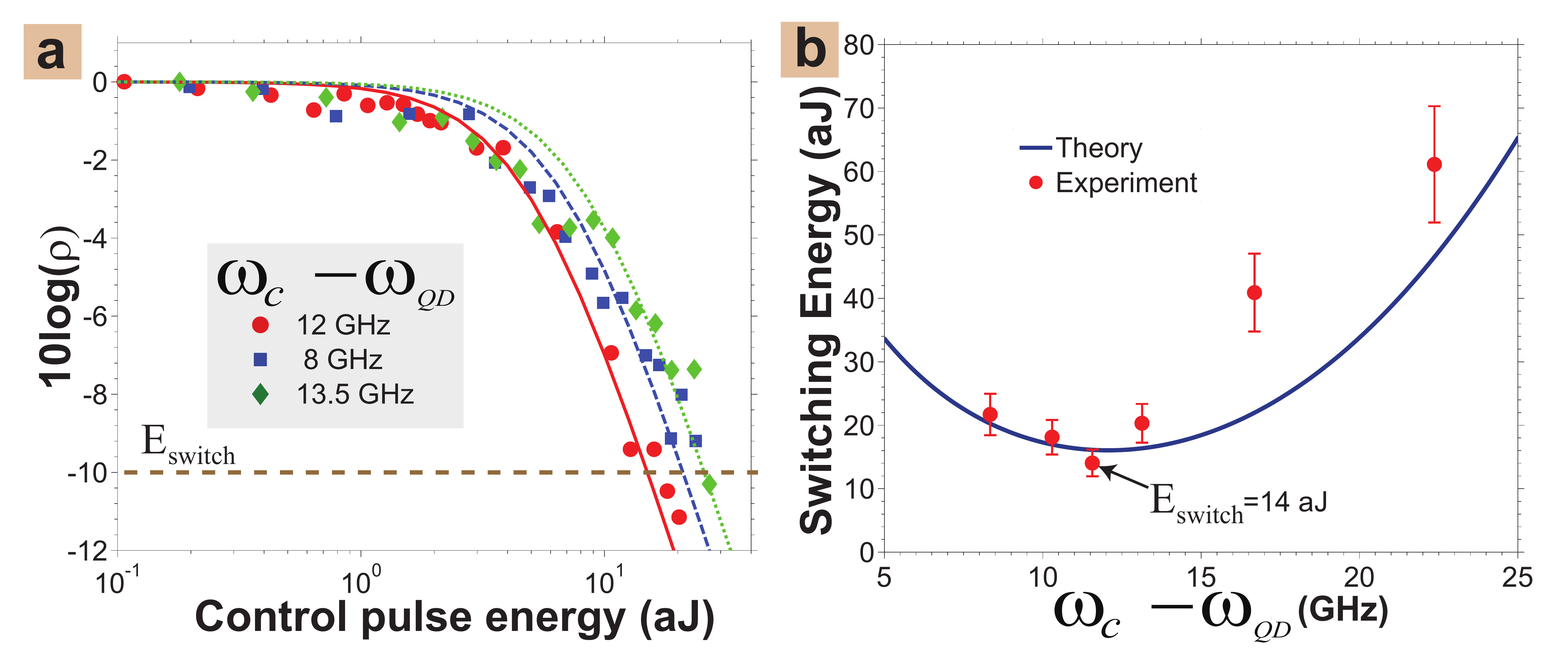}
\caption{(a) Switching contrast   as a function of control pulse energy for three different
detunings between the control pulse and the QD frequency. The solid lines indicate the
theoretical fit, while the horizontal line indicates the 10dB switching point.  (b)
Switching energy as a function of spectral detuning between the control pulse and QD
resonant frequency.  Error bars for the experimental data represent 90\% confidence
intervals for the nonlinear fit.  The solid line indicates theoretical
prediction using a simple Stark shift model.}\label{fig:Pulsed_Energy}
\end{figure}

Having determined the coupling efficiency of the grating coupler into the waveguide, we
can precisely determine the switching energy of the device.  We define the switching
energy $E_{switch}$ of the device as the amount of energy in the control pulse
propagating in the waveguide needed to induce 90\% of the maximum change in reflectivity.
Figure 4a plots the relative change in signal scattering intensity, defined as
$\rho=(\Imax-I)/(\Imax-\Imin)$ where I is the scattering intensity and $\Imax$ and
$\Imin$ are previously defined, as a function of control pulse energy for three different
detunings between the control pulse frequency $\wcon$ and QD frequency $\wqd$. As the
control pulse intensity is increased, the device makes a smooth transition from  $\rho=1$
to an asymptotic value of  $\rho=0$ at high control energies. For each curve, we fit the
data to a theoretical mode given by $\rho(E) = 1/(1+E/E_0)^2$, where $E=\Pwg\eta/R$ is
the control pulse energy, $R=76.3$ MHz is the pulsed laser repetition rate, and $E_0$ is
treated as a fitting parameter. The horizontal line represents 10 dB (90\%) change in
$\rho$, which defines the switching energy. At 12 GHz detuning (red circles), which is
close to resonance with the lower polariton, we achieve a switching energy of
$E_{switch}= 14$ aJ. The blue squares and green diamonds plot the cases where the pump is
blue shifted and red shifted from the lower polariton. In both cases a higher pumping
intensity is required, which manifests itself in a shift of the switching curve to higher
energies. Figure 4b plots the switching energy as a function of detuning of the control
frequency from the QD frequency.  A solid line plots the theoretical curve based on a
semiclassical Stark shift model for the nonlinearity (see Supplementary Information).  A
minimum switching energy is predicted near the lower polariton frequency at $13.4$~GHz
detuning, which is consistent with experimental measurements. When the control laser is
resonant with the lower polariton, the fraction of light coupled to the cavity is given
by $\delta=(1-(1-2\kin/\kappa)^2)=0.36$.  The energy dissipation of the device can be
upper bounded by fraction of control energy that couples to the cavity and is therefore
given by $E_{dis}  = \delta E_{switch} = 5$ attojoules.


In conclusion, we have investigated the dynamic switching properties of a QD strongly
coupled to a photonic crystal cavity. Extremely low switching energies of 14 attojoules
were attained, which are promising numbers for potential applications in all-optical data
routing and optical logic.  Further improvements in the coupling between the cavity and
waveguide could potentially enable optical switching at the single photon level, which is
of great importance in quantum optics and quantum information.

D. Sridharan and R. Bose contributed equally to this work.  This work was supported by a
DARPA Defense Science Office grant (Grant W31P4Q0910013), the Physics Frontier Center at
the Joint Quantum Institute, the Office of Naval Research Applied Electromagnetics
Center, the Army Research Office MURI on hybrid quantum interactions (grant number
W911NF09104), and a National Science Foundation CAREER award (grant number ECCS -
0846494)


\pagebreak

\end{document}